\documentclass[a4paper,preprint,aps,superscriptaddress,nofootinbib]{revtex4-1}
\usepackage[latin9]{inputenc}
\usepackage{color}
\usepackage{amsmath}
\usepackage{amssymb}
\usepackage{graphicx}
\usepackage[unicode=true,pdfusetitle,
 bookmarks=true,bookmarksnumbered=false,bookmarksopen=false,
 breaklinks=false,pdfborder={0 0 1},backref=false,colorlinks=true]
 {hyperref}
\hypersetup{
 citecolor=blue}

\makeatletter

\usepackage[normalem]{ulem}
\usepackage{dcolumn}
\usepackage{bm}
\usepackage{braket}
\usepackage{slashed}

\usepackage{subcaption}
\usepackage{float}
\usepackage{indentfirst}
\usepackage{adjustbox}
\usepackage{tabularx}

\newcommand{\Blue}[1]{\textcolor{blue}{#1}}

\makeatother

\begin{document}
\title{Discovery potential of multi-ton xenon detectors in neutrino electromagnetic
properties}
\author{Chung-Chun Hsieh}
\affiliation{Department of Physics, Center for Theoretical Physics, and Leung Center
for Cosmology and Particle Astrophysics, National Taiwan University,
Taipei, Taiwan 106}
\author{Lakhwinder~Singh}
\affiliation{Institute of Physics, Academia Sinica, Taipei 11529, Taiwan}
\affiliation{Department of Physics, Banaras Hindu University, Varanasi 221005,
India}
\author{Chih-Pan~Wu}
\affiliation{Department of Physics, Center for Theoretical Physics, and Leung Center
for Cosmology and Particle Astrophysics, National Taiwan University,
Taipei, Taiwan 106}
\author{Jiunn-Wei~Chen\footnote{E-mail: jwc@phys.ntu.edu.tw}}
\affiliation{Department of Physics, Center for Theoretical Physics, and Leung Center
for Cosmology and Particle Astrophysics, National Taiwan University,
Taipei, Taiwan 106}
\affiliation{Center for Theoretical Physics, Massachusetts Institute of Technology,
Cambridge, MA 02139, USA}
\author{Hsin-Chang~Chi}
\affiliation{Department of Physics, National Dong Hwa University, Shoufeng, Hualien
97401, Taiwan}
\author{C.-P.~Liu\footnote{E-mail: cpliu@mail.ndhu.edu.tw}}
\affiliation{Department of Physics, National Dong Hwa University, Shoufeng, Hualien
97401, Taiwan}
\author{Mukesh K. Pandey}
\affiliation{Department of Physics, Center for Theoretical Physics, and Leung Center
for Cosmology and Particle Astrophysics, National Taiwan University,
Taipei, Taiwan 106}
\author{Henry~T.~Wong}
\affiliation{Institute of Physics, Academia Sinica, Taipei 11529, Taiwan}
\date{\today}
\begin{abstract}

Next-generation xenon detectors with multi-ton-year exposure are
powerful direct probes of dark matter candidates,
in particular the favorite weakly-interacting massive particles.
Coupled with the features of low thresholds and backgrounds,
they are also excellent telescopes of solar neutrinos. In this paper,
we study the discovery potential of ton-scale xenon detectors
in electromagnetic moments of solar neutrinos. Relevant neutrino-atom
scattering processes are calculated by applying a
state-of-the-arts atomic many-body method---relativistic random phase
approximation (RRPA). Limits on these moments are
derived from existing data and estimated with future
experiment specifications. With one ton-year exposure, XENON-1T can improve the effective milli-charge constraint by a factor two. With LZ and DARWIN, the projected improvement on the solar neutrino effective milli-charge(magnetic moment) is around 7(2) times smaller than the current bound.
 If LZ can keep the same background level and push the electron recoil threshold to 0.5 keV, the projected improvement on milli-charge(magnetic moment) is about $10(3)$ times smaller than the current bound.

\end{abstract}
\maketitle

\section{Introduction}

Xenon detectors play a dominant role in direct experimental searches of a favorite dark matter candidate---weakly-interacting massive particles (WIMPs). The current best limits on the spin-independent WIMP-nucleon cross section are set by xenon detectors with
$4.1\times10^{-47}\,\textrm{cm}^{2}$ for a $30\,\textrm{GeV}/c^{2}$
WIMP and $8.6\times10^{-47}\,\textrm{cm}^{2}$ for a $50\,\textrm{GeV}/c^{2}$
WIMP set by XENON1T~\citep{WIMP-XENON1T} and
PandaX-II~\citep{WIMP-PandaX-II} with sub-ton-year exposure. It is expected that the next-generation
experiments, XENONnT~\citep{G2Xe-XENON1T}, LZ~\citep{G2Xe-LZ}, and DARWIN~\citep{G2Xe-DARWIN}, with multi-ton-year exposure will bring
further improvement by one or two orders of magnitude.

To achieve extremely low background, WIMP detectors are typically hosted in deep underground laboratories to shield cosmic rays. However, neutrinos
from the Sun, supernovae and their remnants, and atmospheric showers can still reach the detectors to generate scattering events indistinguishable from the WIMP signals and form the so-called irreducible background. In the sub-keV to 100 keV recoil energy that WIMP searches focus on, neutrino-nucleus coherent scattering and neutrino induced atomic ionization are the main sources of background and are studied in Refs.~\citep{nubg_NR} and \citep{nubg_ER-FEA,nubg_ER-RRPA}, respectively.

This feature implies that those ton-scale, low background xenon WIMP detectors can serve as excellent solar neutrino telescopes to study solar and neutrino physics at the same time. For example, the very low energy solar neutrino flux from proton-proton fusion can be measured to the $1\%$-level precision \citep{nubg_ER-FEA}, which provides experimental check to the Standard Solar Model. The high energy solar neutrino flux from the $^{8}\textrm{B}$ decays can trigger observable coherent neutrino-nucleus scattering. 

In this paper, we attempt to address the discovery potential of using ton-scale liquid xenon detectors to probe exotic electromagnetic (EM) properties of
solar neutrinos. In the Standard Model (SM) of particle physics, neutrinos are charge neutral and have extremely tiny EM moments including magnetic
dipoles, electric dipoles, anapoles, and charge radii through radiative corrections. Those SM EM properties are not detectable with current experimental sensitivity but new physics beyond SM might make them detectable. Hence the detection of those exotic EM properties are important probes of new physics which could have profound implications to particle physics and astrophysics (see, e.g., Refs.~\citep{PDG,nuEM_rev} for recent reviews of this topic). 

The current best limits reported by direct laboratory searches for
``effective'' (to be explained in the next section) neutrino EM
moments include the following: the magnetic moment of reactor anti-electron
neutrinos $\mu_{\bar{\nu}_{e}}^{\textrm{eff}}<2.9\times10^{-11}\mu_{\textrm{B}}$
($\mu_{\textrm{B}}$ is the Bohr magneton) by the GEMMA experiment
using a germanium detector~\citep{best-MM-GEMMA}; the magnetic moment
of solar neutrinos $\mu_{\nu_{\textrm{S}}}^{\textrm{eff}}<2.8\times10^{-11}\mu_{\textrm{B}}$
by the Borexino experiment using a liquid scintillator~\citep{best-MM-Borexino};
the effective milli-charge of reactor anti-electron neutrinos $\delta_{\bar{\nu}_{e}}^{\textrm{eff}}<1.5\times10^{-12}e_{0}$
($e_{0}$ is the positron charge) using data from the GEMMA experiment~\citep{best-mQ-GEMMA},
and  $\delta_{\bar{\nu}_{e}}^{\textrm{eff}}<2.1\times10^{-12}e_{0}$
by the TEXONO experiment using a smaller germanium detector with a
lower-threshold than the one of GEMMA~\citep{best-mQ-TEXONO}; and
the effective charge radius squared of reactor anti-electron neutrinos
$\braket{r_{\bar{\nu}_{e}}^{2}}^{\textrm{eff}}<3.3\times10^{-32}\,\textrm{cm}^{2}$
by the TEXONO experiment using a CsI scintillator~\citep{best-r2}.
Our main goal of this study is to quantify whether these limits can
be improved with xenon detectors with multi-ton-year exposure to solar
neutrinos using realistic detector specifications of energy threshold,
energy resolution, and background.

To achieve this goal, we need to calculate the scattering cross sections
of solar neutrinos and xenon atoms, and use them to predict the event
rates expected at detectors. Because the energy scales in these scattering
processes, which range from a few tens of eV to a few tens of keV,
overlap with the ones of atomic physics, it is necessary to perform
reliable many-body computations. Similar to our previous works on
neutrino-germanium scattering~\citep{Chen-MM,best-mQ-TEXONO,Chen-EM},
we apply an \textit{ab initio }approach, the relativistic random phase
approximation (RRPA)~\citep{RRPA1,RRPA2,RRPA3,RRPA4}, to neutrino-xenon
scattering. We report new results of neutrino-Xenon scattering through exotic EM interactions, together with the weak interaction process 
calculated in Ref.~\citep{nubg_ER-RRPA} for completeness.

\section{Scattering of Solar Neutrinos off Xenon Atoms}

In the most general case, the EM current of a neutrino field, $\nu$,
is given by 
\begin{align}
j_{\mu}^{(\gamma)}= & \bar{\nu}\left[F_{1}(q^{2})\gamma_{\mu}-i(F_{2}(q^{2})+iF_{E}(q^{2})\gamma_{5})\sigma_{\mu\nu}q^{\nu}+F_{A}(q^{2})(q^{2}\gamma_{\mu}-\slashed{q}q_{\mu})\gamma_{5}\right]\nu\,,\label{eq:j_nu_EM}
\end{align}
where $q_{\mu}$ is the four momentum transfer, $q^{2}=q_{\mu}q^{\mu}$,
$\slashed{q}=q_{\mu}\gamma^{\mu}$, and $\gamma_{\mu}$, $\gamma_{5}$,
and $\sigma_{\mu\nu}$ the standard Dirac matrices. The formfactors near the $q^{2}\rightarrow0$ limit yield the definitions of 
milli-charge $\delta_{Q}=F_{1}(0)$, magnetic moment $\mu_{\nu}=F_{2}(0)$, electric
dipole moment $d_{\nu}=F_{E}(0)$, and anapole moment $a_{\nu}=F_{A}(0)$ of a neutrino. 

There are some important theory aspects to point out here: First,
consider neutrino scattering off a free electron, the differential
cross section $d\sigma/dT$ with respect to neutrino energy deposition
$T$, scales as $T^{-2}$ for the milli-charge interaction~\citep{nu-e_mQ-1,best-mQ-GEMMA}, $T^{-1}$ for the magnetic and
electric dipole moment interaction~\citep{nu-e_MM-1,nu-e_MM-2},
and $T^{0}$~\citep{nu-e_r2} for the anapole moment and charge-radius-squared interaction.
This implies enhanced sensitivities to
$\delta_{Q}$, $\mu_{\nu}$ and $d_{\nu}$ for detectors with low
thresholds. Also note that the interactions with $a_{\nu}$ and $\braket{r_{\nu}^{2}}$
have the same contact forms as the low-energy weak interactions, so
they can be effectively included by modifying the neutrino weak coupling
strengths.

Second, as solar neutrinos are ultra-relativistic, that is, $E_{\nu}\gg m_{\nu}$,
it is a good approximation to set $m_{\nu}=0$ in scattering. At this
limit, a neutrino helicity eigenstate is also a chirality eigenstate.
By the identities 
\begin{equation}
\bar{\nu}_{R}\sigma_{\mu\nu}\nu_{L}=-\bar{\nu}_{R}\sigma_{\mu\nu}\gamma_{5}\nu_{L}\,,\quad\bar{\nu}_{L}\gamma_{\mu}\nu_{L}=-\bar{\nu}_{L}\gamma_{\mu}\gamma_{5}\nu_{L}\,,
\end{equation}
one sees that the $\mu_{\nu}$ and $d_{\nu}$ interactions are not distinguishable, nor are the $a_{\nu}$ and $\braket{r_{\nu}^{2}}$
interactions. They will appear in scattering amplitudes as linear
combinations $\mu_{\nu}-id_{\nu}$ and $\braket{r_{\nu}^{2}}-6a_{\nu}$.

Third, because neutrinos oscillate and the final neutrino type is
not observed in the detector setup that we are discussing, the general
EM moments of neutrinos should be written in a $3\times3$ matrix form
to accommodate the possible transition EM moments. The matrixes of the moments  involve neutrinos
of different types in the incoming and outgoing states, in addition
to the static EM moments, which are the diagonal matrix elements.
Thus for solar neutrinos, the magnetic moment probed at the detector
end is actually an ``effective'' one which appears in the cross section
in a squared form as 
\begin{equation}
\left|\mu_{\nu_{\textrm{S}}}^{\mathrm{eff}}\right|^{2}=\sum_{f}\left|\sum_{i}A_{ie}(E_{\nu},L)\left(\mu_{fi}-id_{fi}\right)\right|^{2}\,,
\end{equation}
where $f$ and $i$ are the mass eigenstate indices for the outgoing
and incoming neutrinos at the point of the neutrino EM interaction
caused by the transition magnetic and electric dipole moments $\mu_{fi}$
and $d_{fi}$, respectively~\citep{mu_nu_eff-1,nuEM_rev,best-MM-Borexino}. The
amplitude $A_{ie}(E_{\nu},L)$ describes how a solar neutrino, which
is an electron neutrino $\nu_{e}$ at birth, oscillates to a mass
eigenstate $\nu_{i}$ through the in-medium oscillation in the solar
interior and the subsequent vacuum oscillation while traversing from
the surface of the Sun to the Earth. The summation over $i$ gives
the total transition amplitude (squared to give probability), and
the summation over $f$ indicates the incoherent sum over different final
states. Similar procedures should be applied to effective
milli-charge $\delta_{Q_{\textrm{S}}}^{\textrm{eff}}$ and charge
radius squared $\braket{r_{\nu_{\textrm{S}}}^{2}}^{\textrm{eff}}$.
Note that these effective moments depend on neutrino energy $E_{\nu}$
and the Sun-Earth distance $L$. But in practice, they can be approximated
as constants~\citep{nuEM_rev,best-MM-Borexino}. In this work, we
only concern how effective moments are constrained by experiments
but not their composition and their dependence on neutrino oscillation
parameters.

The majority of solar neutrinos comes from two sources which make up about $98\%$ of the total solar neutrino flux: the proton-proton
(pp) fusion, $p+p\rightarrow d+e^{+}+\nu_{e}$, which produces a continuous
spectrum with $E_{\nu}$ from 0 to 420 keV~\citep{Solar3}, and the
electron capture by $^{7}$Be, $^{7}\textrm{Be}+e^{-}\rightarrow{}^{7}\textrm{Li}+\nu_{e}$,
which produces two discrete spectral lines at 862 keV and 384 keV
with branching ratios $89.6\%$ and $10.4\%$. Their fluxes on the Earth surface are~\citep{Solar1,Solar2}
\begin{equation}
\phi_{\textrm{pp}}=5.98\times10^{10}\,\textrm{cm}^{-2}\,\textrm{s}^{-1},\;\phi_{^{7}\textrm{Be}}=5.00\times10^{9}\,\textrm{cm}^{-2}\,\textrm{s}^{-1}\,.
\end{equation}
Compared with experiments using germanium detectors to probe the EM
moments of reactor antineutrinos, the solar neutrino flux is smaller
by 2-3 orders of magnitude. However, ton-scale xenon detectors can
make up this deficiency by target mass, as typical germanium
detectors only operate at the kg-scale.

Both pp and $^{7}\textrm{Be}$ neutrinos have rather low energies.
They can not deposit observable energy at the keV or sub-keV scale
by scattering off the whole atom or the atomic nucleus elastically.
Therefore, the most effective channel to probe the EM moments of solar
neutrinos is to ionize the xenon atom
\begin{equation}
\nu_{\textrm{S}}(\textrm{pp},^{7}\textrm{Be})+\textrm{Xe}\xrightarrow[(\textrm{weak})]{\textrm{EM}}\nu_{\textrm{S}}+\textrm{Xe}^{+}+e^{-}\,,\label{eq:AI}
\end{equation}
and produces an electron recoil (ER) with a few tens of eV to a few tens of keV in energy. Apparently, the weak scattering is the main background thats limit the discovery potential.
However, xenon detectors can not differentiate an ER event from a nuclear recoil (NR) event below certain energy, currently around 1.4 keV~\cite{WIMP-XENON1T}.\footnote{Ref.~\cite{WIMP-XENON1T} assigns thresholds of 1.4 keV$_{ER}$ and 4.9 keV$_{NR}$ to detect ER and NR {\em signals}, respectively. The NR threshold is higher because of quenching effect.
}
Therefore,  NR events dominantly by coherent neutrino-nucleus scattering (CNNS)
  by the $^{8}\textrm{B}$ neutrinos 
\begin{equation}
\nu_{\textrm{S}}(^{8}\textrm{B})+\textrm{Xe}\xrightarrow[(\textrm{CNNS})]{}\nu_{\textrm{S}}+\textrm{Xe}\,,\label{eq:CNNS}
\end{equation}
in the range $\sim20\,\textrm{eV}$-$2\,\textrm{keV}$, should also be considered as a background.

The formalism for calculating the differential cross sections of neutrino-atom
ionization of Eq.~(\ref{eq:AI}) is documented in detail in Ref.~\citep{Chen-EM}. The
formalism for CNNS of Eq.~(\ref{eq:CNNS}) can be found in
Ref.~\citep{CNNS}. We only re-iterate the importance of atomic structure in these calculations, and highlight
some key features of our results. 

A neutral xenon atom has $Z=54$ electrons, so the relativistic correction
can be as large as $Z\alpha\sim0.4$. The leading relativistic correction
and two-electron correlations are taken into account in our calculation
through RRPA~\citep{RRPA1,RRPA2,RRPA3,RRPA4}. This approach was
benchmarked by the photo-xenon absorption process shown in Fig.~1 of Ref.~\citep{nubg_ER-RRPA}.
The calculation
agrees with the experimental data~\citep{data1,data2,data3,data4,data5}
at a few percent level between 100 eV to 30 keV, except near atomic
ionization thresholds or ``edges.'' A photoabsorption edge is where the cross section
reaches a local maximum. So a small mismatch between the experimentally-measured
and the theoretically-calculated energy edges leads to a large cross
section difference around it. However, this mismatch is largely removed by considering the finite
energy resolution of a detector. The
theory error is conservatively estimated to be 5\% in this range.
Between 70-100 eV, the large peak at around 100 eV indicates strong
correlations. Due to slow convergence of the computation in this range, only a
few points were obtained. The largest disagreement is seen between
40-70 eV but still within $30$\%.


\begin{figure}[t]
	\includegraphics[width=0.7\linewidth]{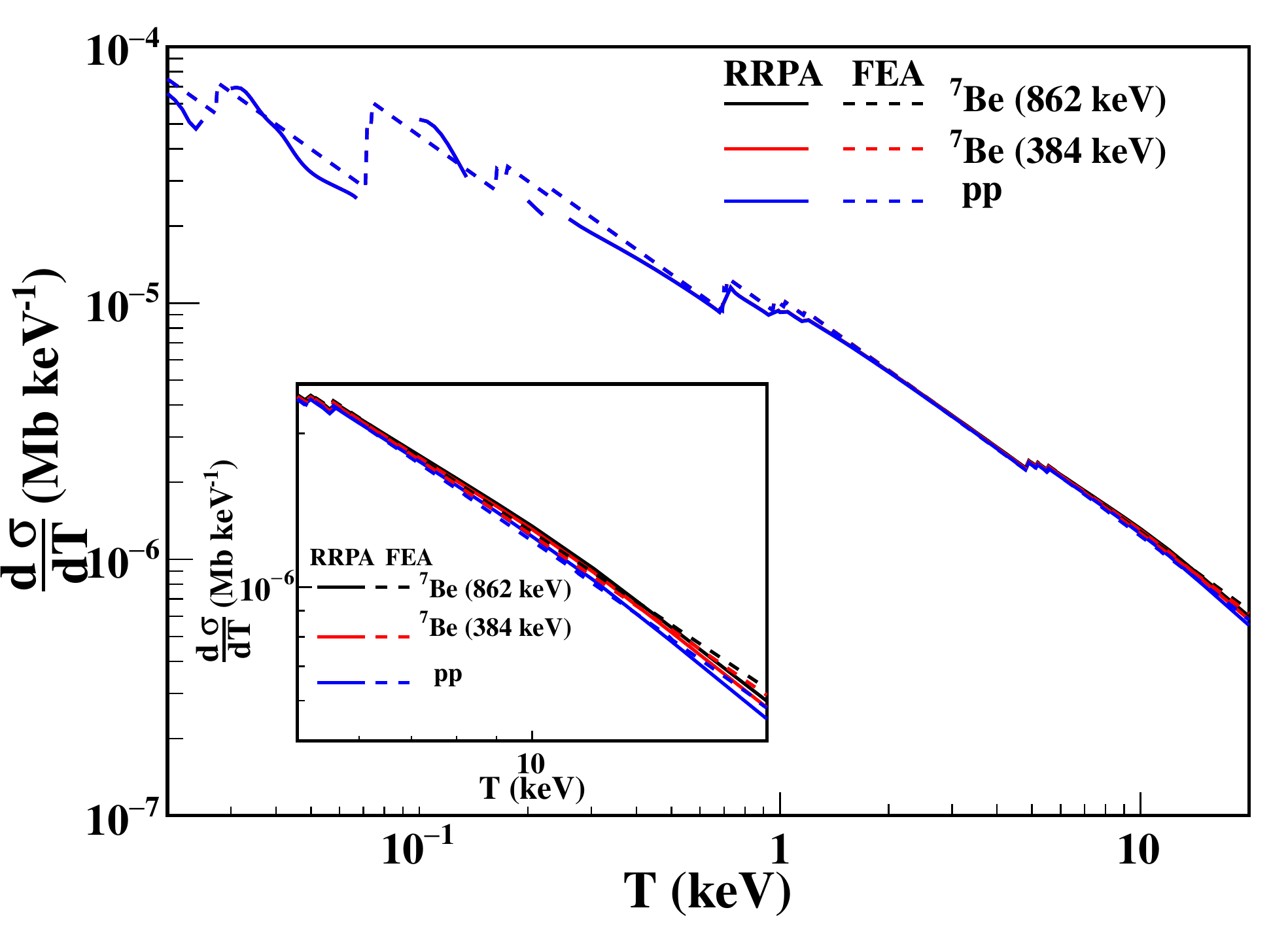}
	\caption{$\nu$-Xe differential cross sections through the neutrino magnetic moment
		interaction for the $^{7}$Be (862 keV, in black), $^{7}$Be (384
		keV, in blue) and flux averaged pp (in red) solar neutrinos, respectively.
		These curves largely overlap below 5 keV. The difference is highlighted in the inset.
		The dashed curves are the results of FEA.  \label{fig:NMM}}
\end{figure}

In Fig.~\ref{fig:NMM}, the $\nu$-Xe differential cross sections through the neutrino magnetic moment interaction are shown. Contributions from
the $^{7}$Be line spectrum at 862 keV and 384 keV, and the flux averaged
pp neutrinos all have very similar $T$ dependence. We also show the
result using the stepping free electron approximation (FEA): 
\begin{equation}
\frac{d\sigma^{(i)}}{dT}=\sum_{i=1}^{Z}\theta(T-B_{i})\frac{d\sigma_{0}^{(i)}}{dT}\,.
\end{equation}
This is done by weighting the scattering cross section of a neutrino
and a free electron, $d\sigma_{0}/dT$, with the number of electrons
that can be ionized by an energy deposition of $T$, with $\theta$
the step function and $B_{i}$ the binding energy of the $i$th
electron. FEA is shown to be a good approximation for this process
from a sum rule analysis~\citep{Russian1,Russian2,Russian3}. This conclusion was
confirmed numerically by the \emph{ab initio} calculation with Ge in Ref.~\citep{Chen-MM}
and with Xe in this work. However, the \emph{ab
initio} result is consistently smaller than that of FEA in Ge but
not in Xe here. This can be traced back to the approximation
used in the sum rule method which keeps only the leading single electron
operator $J^{0}=e^{\dagger}e$ in the non-relativistic expansion of
the electromagnetic current. The $J^{i}$ current ($i=1,2,3$), which
is suppressed by the electron velocity, and higher dimensional operators
such as $\delta J^{0}=(e^{\dagger}e)^{2}$, are not included in the
analysis. However, from an effective field theory point of view, all operators
of the same quantum numbers as $J^{0}$ and $J^{i}$ can appear after the high
energy mode mode in the non-relativistic effective field theory is ``integrated out.'' For
example, the one electron operator $e^{\dagger}e$ can exchange one
high energy photon with another electron. After integrating out the high energy photon, the $\delta J^{0}$
operator appears. With just the $J^{0}$ operator, one can show that FEA
yields an upper bound. However, the inclusion of $\delta J^{0}$
could change this conclusion. A similar conclusion was reached in
Ref.~\citep{Russian3} that electron correlations could modify the sum rule analysis of~\citep{Russian1}.
Indeed, our result becomes bigger than FEA near 100 eV where
a large peak in photo-absorption is observed in Fig.~\ref{fig:NMM} of Ref.~\citep{nubg_ER-RRPA}.
This peak should come from large electron correlations since the nearest
threshold is still $30$~eV away.


\begin{figure}[t]
\includegraphics[width=0.9\linewidth]{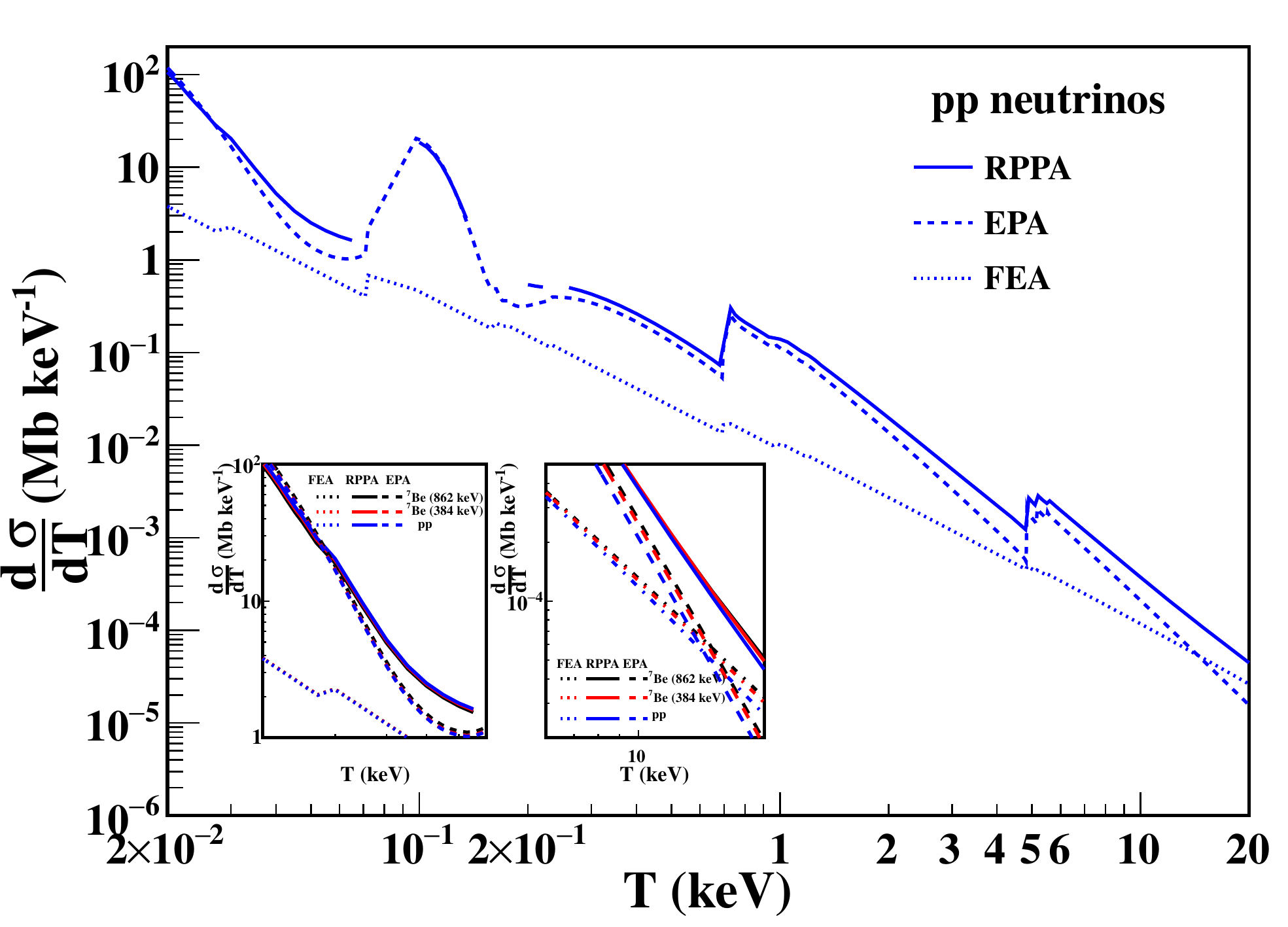}
\caption{$\nu$-Xe differential cross sections through
the neutrino milli-charge interaction
for the $^{7}$Be (862 keV, in black), $^{7}$Be (384 keV, in blue)
and flux averaged pp (in red) solar neutrinos, respectively. 
These curves largely overlap. The difference is highlighted in the insets.
The FEA and EPA results are also shown. \label{fig:mQ} }
\end{figure}

Analogously, the $\nu$-Xe differential cross sections through the neutrino milli-charge
interaction are shown in Fig.~\ref{fig:mQ}. The milli-charge interaction has a sharper $T$ dependence than the neutrino magnetic moment interaction
which is also seen in scattering over a free electron. 
It has been shown in the previous study that at low energies, the equivalent photon approximation
(EPA), which relates the ionization cross section to the one of photo-absorption,
works well in the Ge ionization by the neutrino milli-charge interaction~\citep{best-mQ-TEXONO}.
While at energies higher than all the electron binding energies, electron
binding is no longer important and the result approaches the EFA curve~\citep{best-mQ-TEXONO}.
These features are also seen in the Xe case here.

\section{Results and Discussions}

The differential count rates as functions of neutrino energy deposition,
$dR(T)/dT$, are obtained by convoluting the differential cross sections
$d\sigma(T,E_{\nu})/dT$ with the incident solar neutrino energy spectrum
$d\phi(E_{\nu})/dE_{\nu}$ 
\begin{equation}
\frac{dR(T)}{dT}=\frac{N_{0}\Delta t}{A}\int dE_{\nu}\frac{d\sigma(T,E_{\nu})}{dT}\frac{d\phi(E_{\nu})}{dE_{\nu}} ,
\end{equation}
where the unit is fixed as events per ton per year per keV, $N_{0}=6.02\times10^{29}$,
 $A$ is the atomic mass of the atom, and $\Delta t=1\,\textrm{year}$.

\begin{figure}[t]
\includegraphics[width=0.8\linewidth]{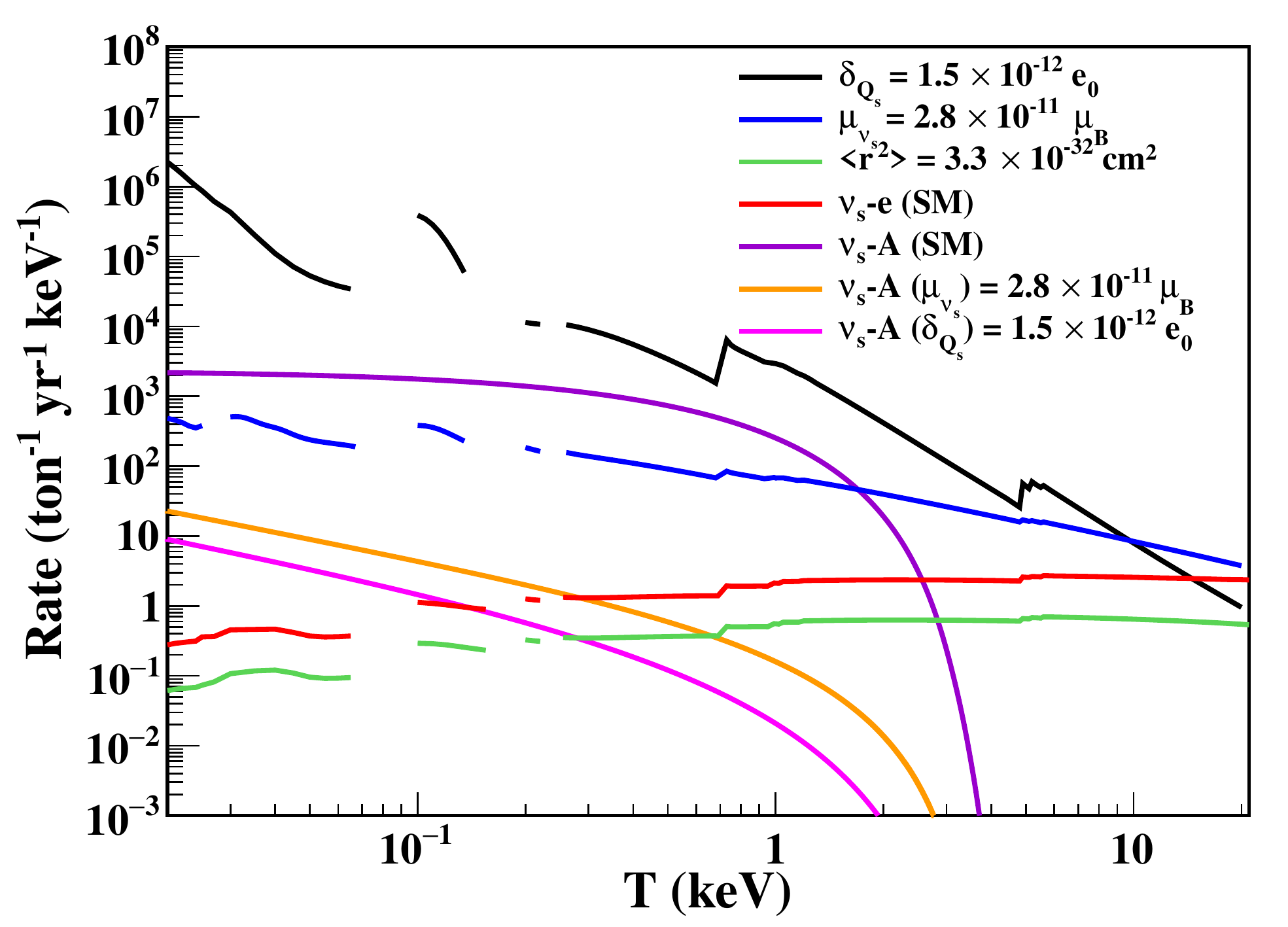}
 \caption{ Electron and nuclear recoil event rates from solar neutrino Xenon scattering. 
 Electron recoil channels: neutrino atomic ionization through weak interaction
(red), neutrino milli-charge (black, scales as $\delta_{Q_{S}}^{2}$), magnetic moment (blue, scales as $\mu_{\nu_{S}}^{2}$), and charge
radius (green, scales as $<r^{2}>$, from interference with the weak interaction contribution). Nuclear recoil channels: neutrino nucleus coherent scatterings through weak interaction
(purple) neutrino magnetic moment (golden, scales as $\mu_{\nu_{S}}^{2}$), and neutrino milli-charge (magenta, scales as $\delta_{Q_{S}}$, from interference with the weak interaction contribution). 
$e_{0}$ is the electron charge and $\mu_{B}$ is Bohr magneton.
Some lines are broken due to slow convergence of the RRPA calculation in the broken areas. 
\label{fig:money}}
\end{figure}

We have shown relevant solar neutrino xenon scattering processes 
using the current best upper limits on neutrino EM moments as inputs for the effective EM moments of solar neutrinos in Fig.~\ref{fig:money}: $\delta_{Q_{\textrm{S}}}^{\textrm{eff}}=1.5\times10^{-12}\,e_{0}$, $\mu_{\nu_{\textrm{S}}}^{\textrm{eff}}=2.8\times10^{-11}\,\mu_{\textrm{B}}$, and $\braket{r_{\nu_{\textrm{S}}}^{2}}^{\textrm{eff}}= 3.3\times10^{-32}\,\textrm{cm}^{2}$. The rates are with one ton-year exposure.

The red curve is for the electron recoil from atomic ionization ($\nu+A\to\nu+A^{+}+e^{-}$)
through the standard weak scattering which is considered as a background in this work. In the range of $T$  considered,
major contributions are from the pp and $^{7}\textrm{Be}$ neutrinos.
Note that charged-current interaction is not flavor-blind ($\sigma_{\nu_{\mu}}=\sigma_{\nu_{\tau}}\ne\sigma_{\nu_{e}}$)
and will depend on solar neutrino oscillations. The result presented here
is based on the same computation scheme as the one of Ref.~\citep{nubg_ER-RRPA} with neutrino oscillations included
but with improved error estimation. The differential cross section
at large $T$ is slightly increased, but still smaller than the calculation~{\citep{nubg_ER-FEA}}
based on free electron scattering by 19\%.

The purple curve is for the nuclear recoil
from neutrino-nucleus coherent
scattering ($\nu+A\to\nu+A$) which is also considered as a background here since ER and NR events cannot be distinguished at low energies (current below 1.4 keV). At $T>0.01\,\textrm{keV$_{NR}$}$, it is dominated
by the $^{8}\textrm{B}$ neutrinos, because pp and $^{7}\textrm{Be}$ neutrinos
are not energetic enough. 
The process only involves the flavor-blind neutral-current interaction so it is independent of
solar neutrino oscillations. The event rate presented here is based on
the standard formula for neutrino-nucleus coherent scattering,
quenching factor Q$_{f}$ = 1 ({\em i.e.} no quenching), and the coherency is more than 0.9 in the range of $T$~\citep{CNNS}.




$\nu$-Xe EM ionization event rates are presented by black (milli-charge),
blue(magnetic moment), and green curves (charge radius squared), respectively. 
The result suggests that if the detector has good sensitivity to the weak interaction (red curve) background then the current bound on neutrino effective milli-charge can be further improved. The improvement on the current bound of neutrino effective magnetic moment is also possible especially with $T > 2\,\textrm{keV}$. Below 2 keV the NR background from neutrino-nucleus coherent scattering (purple curve) needs to be subtracted which might be a challenge.

We also study the possible signal of $\nu$-Xe EM interaction through the NR process. The golden curve is neutrino-nucleus coherent scattering through neutrino magnetic moment using
\begin{equation}
  \frac{d \sigma_{\mu} (\nu A \rightarrow\nu A )}{dT_{NR}} = \frac{\pi \mu^{2}_{\nu} \alpha^{2} Z^{2}}{m_{e}^{2}} \left(\frac{1}{T_{NR}} -\frac{1}{E_{\nu}} +  \frac{T_{NR}}{4E_{\nu}^{2}} \right) F_{V}^{2}(q^{2})
\end{equation}
where $Z$ is the  atomic  number,  $m_{e}$ is the electron mass coming from the definition of Bohr magneton,
$F_{V}(q^2)$ with normalization $F_{V}(0)=1$ is  the nuclear isoscalar vector form factor for xenon which is an isoscalar neucleus, and the on-shell condition for the nucleus fixes $q^2 = -2 M_A T_{NR}$~\cite{nu-e_MM-2}. 
The effect is much smaller than coherent scattering through weak interaction and is also much smaller than ER through neutrino magnetic moment interaction.  Similar NR process through neutrino milli-charge can interfere with the weak interaction process:
\begin{eqnarray}
  \frac{d\sigma_{\delta_{Q}}(\nu A \rightarrow\nu A )}{dT_{NR}} & = & \frac{M_A G_F^2}{4 \pi}(1-\frac{M_{A}T_{NR}}{2 E_{\nu}^{2}})( 2 A \sin^2 \theta_W+x)^2 F_{V}^{2}(q^{2}) , \nonumber \\
  x & = & \frac{2 \sqrt{2} \pi \alpha Z \delta_{Q}}{G_{F}M_{A }T_{NR}} ,
  \label{EQ:nN-mQ}
\end{eqnarray}
where $G_{F}$ is the Fermi coupling constant, $\theta_{W}$ is the Weinberg angel, and $M_{A}$ is the nuclear mass. The current $\delta_{Q}$ bound makes the interference term (the $x$ term) much bigger than the $x^{2}$, hence we present the $x$ term as  the magenta curve.

To put the above naive (and optimistic, too) prospect on a more realistic
ground, one needs to take detector specifications into account. 
All relevant backgrounds can be categorized into external and  intrinsic ones.
The external background comes from radioactive contamination in detector construction materials which can be reduced considerably by target fiducialization. 
However, this self-shielding does not work for the intrinsic background from $^{85}$Kr,  $^{222}$Rn,
and  double beta  decays  of $^{136}$Xe. A potential background for neutrino-electron interactions comes from
these components. Based on earlier studies~\cite{Darwin-proj,multi-ton-LXe}, the total ER background
distribution is assumed to be flat below 20 keV. The upper limits on the various neutrino electromagnetic
parameters are derived under the assumption that the predicted signal cannot exceed the measured
background rate. The energy resolution for xenon detectors is adopted from Ref.~\Blue{\cite{xe100-reso}}. 

It is also worth emphasizing that the distinguishing power of NR and ER for
liquid Xe detectors relies on the S1 (prompt scintillation) signal. In the absence of S1 signal,
the electromagnetic background contamination increases significantly due
to losing $z$-coordinate reconstruction from the time difference of S2/S1  and
particle identification based on the S2/S1 signal ratio. 
More specifically, although the S2 energy threshold is lower than S1 and can be pushed to sub-keV due to electroluminescence amplification, also
the ambient background can be minimized by using a smaller fiducial mass, however, the NR background from neutrino-nucleus coherent scattering becomes a major issue in the absence of discrimination between NR and ER events in the S2-only analysis.  Therefore, we will only present constraints from the S1+S2 combined analyses.

In Table~\ref{tab::results}, we show the key detector parameters for LXe detectors and their sensitivity on various solar neutrino EM moments
for existing  experiments: XENON-10~\citep{xe10-data}, XENON-100~\citep{xe100-data}, PandaX-II~\citep{pandaxII-data}, XENON-1T~\citep{xe1t-data}, and future experiments: LZ~\citep{LZ-report} and DARWIN~\citep{Darwin-proj}. It is interesting that with one ton-year exposure, XENON-1T can improve the $\delta_{Q_{\textrm{S}}}^{\textrm{eff}}$ constraint by a factor two. With LZ and DARWIN, the projected improvement on $\delta_{Q_{\textrm{S}}}^{\textrm{eff}}$ is around 7 times smaller than the current bound, while the projected improvement on $\mu_{\nu_{\textrm{S}}}^{\textrm{eff}}$ is around a factor two.
To further motivate future experimental effort, if LZ can keep the same background level and push the ER threshold to 0.5 keV$_{ER}$ with S1+S2,
then limits of  $\delta_{Q_{\textrm{S}}}^{\textrm{eff}}\sim10^{-13} e_{0}$(about one order smaller than the current bound) and $\mu_{\nu_{\textrm{S}}}^{\textrm{eff}} \sim 10^{-11} \mu_{B}$(about  three times smaller than the current bound) can be achieved.

For the effective charge radius squared, we see the constraints in Table~\ref{tab::results} is no better than existing bound. This is because the associated cross section is not enhanced relative to the weak interaction background at low energy. Hence the best constraint is in fact from TEXONO experiment using MeV reactor neutrinos~\cite{best-r2}.

\begin{table*}
\caption{
	Summary of experimental limits at 90\% CL on the effective EM moments:
	$\mu_{\nu_{\textrm{S}}}^{\textrm{eff}}$, $\delta_{Q_{\textrm{S}}}^{\textrm{eff}}$,
	and $\braket{r_{\nu_{\textrm{S}}}^{2}}^{\textrm{eff}}$ of solar neutrinos, assuming an energy resolution
	from the XENON100 experiment~\citep{xe100-reso}.}
\label{tab::results}	
	
	\begin{ruledtabular}
		\begin{tabular}{lcccccc}
			Experiment                         & Exposure               & Threshold          &  Background Level                   &   \multicolumn{3}{c}{Upper Bounds at 90\% CL }  \\
			& (ton-year)             & ($\rm{keV_{ER}}$)  & ($\rm{kg^{-1} keV^{-1} day^{-1}}$)  &  $\mu_{\nu_{\textrm{S}}}^{\textrm{eff}}$                   & $\delta_{Q_{\textrm{S}}}^{\textrm{eff}}$  & $\braket{r_{\nu_{\textrm{S}}}^{2}}^{\textrm{eff}}$      \\
			&                        & (S1 + S2) &                                   &  ($\times 10^{-11}~\mu_{\rm B}$) & ($\times 10^{-12} e_{0}$) & ($\times 10^{-30} ~ \text{cm}^{2}$) \\ \hline
			XENON-10~\cite{xe10-data}          & 8.67 $\times 10^{-2}$  &   2.0              & 1.1                               &  348.74    &  65.45  &  158.86  \\ 
			XENON-100~\cite{xe100-data}        & 2.1 $\times 10^{-2}$   &   5.0              & $\rm{5.3~\times~10^{-3}}$         &  35.13     &  13.03   &  11.20  \\
			PandaX-II~\cite{pandaxII-data}     & 7.4 $\times 10^{-2}$   &  1.2               & $\rm{2.7~\times~10^{-3}}$         &   15.46     &   2.06    &   8.13 \\
			XENON-1T~\cite{xe1t-data}          & 1.0                    &  1.4               & $\rm{2.24~\times~10^{-4}}$        &   4.51      &   0.64   &    2.31  \\
			Projected LZ~\cite{LZ-report}      & 15.34                  & 1.5/0.5                &   $\rm{4.27~\times~10^{-5}}$     &  1.85/$\sim$ 1      &   0.28/$\sim$ 0.01   &    0.93   \\
			Projected DARWIN~\cite{Darwin-proj}& 14.0                   & 2.0                & $\rm{2.0~\times~10^{-5}}$         &   1.27      &   0.24   &  0.58
			
		\end{tabular}
	\end{ruledtabular}

\end{table*}

\section{Summary}

In this work, we study the potential of using current and next-generation
xenon detectors in constraining exotic electromagnetic moments of
neutrinos with low-energy solar pp and $^7\textrm{Be}$ sources. The
cross sections of neutrino-atom scattering due to these electromagnetic neutrino-electron interactions are calculated by the relativistic random phase approximation which is a state-of-the-arts {\it ab initio} approach  
to properly take atomic many-body physics into account. Limits are derived using current experimental data with
sub-ton-year exposure. Projected sensitivity is also estimated
assuming future detector specifications. 

It is interesting that with one ton-year exposure, XENON-1T can improve the effective milli-charge constraint by a factor two. With LZ and DARWIN, the projected improvement on the effective milli-charge is around 7 times smaller than the current bound, while the projected improvement on the effective magnetic moment is around a factor two. 
To further motivate future experimental effort, if LZ can keep the same background level and push the ER threshold to 0.5 keV$_{ER}$ with S1+S2,
then limits of  $\delta_{Q_{\textrm{S}}}^{\textrm{eff}}\sim10^{-13} e_{0}$(one order smaller than the current bound) and $\mu_{\nu_{\textrm{S}}}^{\textrm{eff}} \sim 10^{-11} \mu_{B}$(three times smaller than the current bound) can be achieved.

As for the limit on the effective charge radius of neutrinos, using low energy solar neutrinos have no advantage over using MeV reactor neutrinos. We also consider contributions from additional electromagnetic neutrino-nucleus interactions, but the resulting nuclear recoil signals are also swamped by the background due to coherent neutrino-nucleus scattering.

\begin{acknowledgments}
This work is supported in part by the Ministry of Science and Technology, Taiwan under
Grants Nos. 104-2112-M-259-004-MY3, 105-2112-M-002-017-MY3, 107-2811-M-002-3063, 107-2811-M-002-3097, 107-2112-M-259-002, and 107-2119-M-001-028-MY3; Grant 2019-20/ECP-2 from the National Center for Theoretical Sciences,
and the Kenda foundation.
\end{acknowledgments}

\end{document}